\documentclass[twocolumn,preprintnumbers,amsmath,amssymb]{revtex4}
\usepackage{graphicx}% Include figure files
\usepackage{dcolumn}% Align table columns on decimal point
\usepackage{bm}% bold math

\begin{document}
\title{Graphene and graphane: New stars of nanoscale electronics}
\author{Julia Berashevich and Tapash Chakraborty}
\affiliation{Department of Physics and Astronomy, University of
Manitoba, Winnipeg, MB R3T 2N2, Canada}

\begin{abstract}
Discoveries of graphene and graphane possessing unique electronic
and magnetic properties offer a bright future for carbon based electronics,
with future prospects of superseding silicon in the semiconductor industry.
\end{abstract}

\maketitle

Silicon has been the most widely used material in semiconductor electronics for 
many decades. However, silicon-based devices appear to have several crucial 
disadvantages: low mobility of the charge carriers limits their switching speed 
to the range of a few GHz, the indirect gap renders application of silicon in 
optoelectronics rather inefficient, and finally, silicon technology is ill 
suited for production of nanoscale devices. To overcome these limitations, 
which are becoming crucial with ever diminishing size of the devices, attempts 
to replace silicon by a better semiconductor material have been made with the 
elements of III and V groups of the periodic table (A$^{III}$B$^{V}$), such as 
GaAs. Unfortunately, fabrication of A$^{III}$B$^{V}$-based devices is 
rather expensive, and therefore, their application is justified only for production 
of special optoelectronic devices for which the A$^{III}$B$^{V}$ semiconductors 
with narrow direct gap are most suited. However, even for this application the problem 
of poor compatibility of silicon and A$^{III}$B$^{V}$ technologies in integrated circuits 
has not been resolved satisfactorily. Researchers are therefore desperately looking for 
an alternative material capable of replacing silicon in integrated circuits. While
there are several promising routes that have been explored, carbon electronics seem
to have the brightest future, especially since the discovery of graphene.

\section{Graphene -- A novel carbon material}

Carbon electronics has matured into its own field of reasearch with the discovery of fullerenes 
and carbon nanotubes. These carbon structures have opened a new chapter in condensed-matter 
physics largely due to their unique electronic properties and are also important in
materials science because of their high tensile strength and elasticity. Nevertheless, in 
the past decade, interest in carbon electronics has somewhat faded because of the
difficulties in the development of technology for mass production of three-dimensional 
carbon structures such as fullerenes and nanotubes, and its incompatibility with 
the planar semiconductor technologies. The recent discovery of planar carbon-based materials 
-- graphene and graphane -- is expected to alleviate that latter problem and as a
result, have garnered considerable attention in the communities \cite{julia1,chakra_pic}
as a prospective candidate to replace or to be combined with the silicon technology.

%Figure  1
\begin{figure}
\includegraphics[scale=0.35]{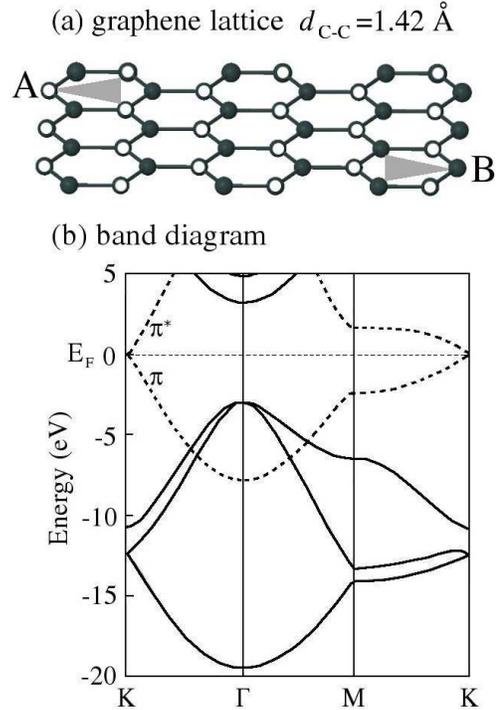}
\caption{\label{fig:fig1} (a) Structure of a single graphene layer (full and empty 
circles correspond to carbon atoms belonging to A and B sublattices, respectively). 
(b) The band diagram of graphene \cite{amara}. The Fermi level ($E_F$) is set to zero energy.}
\end{figure}

Graphene is a single atomic layer of graphite [see the graphene lattice in Fig.~\ref{fig:fig1}(a)]
produced for the first time by mechanical exfoliation of graphite \cite{nov2004}. The carbon atoms in 
graphene are arranged in a hexagonal lattice and are covalently bonded via $sp^2$ hybridization. Three of the four
valence electrons in each carbon atom make three $\sigma$ bonds with its nearest neighbors. 
The covalent $\sigma$ bond provides strong binding between the carbon atoms but the electrons
contribute poorly to the conductivity. One half-filled $p$ orbital left on each carbon 
atom, after its covalent bonding with the neighbors, is orthogonal to the graphene plane. The 
interaction of the $p$ orbitals residing on the nearest-neighboring carbon atoms results in 
the generation of the $\pi$ bonds below and above the graphene plane. The $\pi$ electrons are 
delocalized across the entire lattice and contribute to the conductivity of graphene. 
In the band diagram the $\pi$- electrons are distinguished by the $\pi$ and $\pi^*$ 
bands formed at the top of the valence band and at the bottom of the conduction band, 
respectively, as shown in Fig.~\ref{fig:fig1}(b) \cite{amara}. These $\pi$ bands
meet and become cone shaped at the K-points 
such that they are degenerate and display a linear dispersion near the K-points 
resulting in zero effective mass for electrons and holes and high mobility of the charge 
carriers \cite{julia1}. 

Graphene exhibits many attractive material properties that are important for building electronic
devices. These include high mobilities of charge carriers, high current-carrying
capabilities, high transparency, high thermal conductivity and mechanical stability.
But, despite all these impressive qualities, the absence of a bandgap in graphene,
which renders it a semi-metal, is a major obstacle in developing graphene based electronic
devices. As in all semiconductor systems, the presence of a bandgap is essential for
electronic control of the conductivity. The zero effective mass of electrons and holes in graphene 
creates yet another problem. The charge carriers in graphene can not be confined by the potential 
barriers which is, in fact, transparent for these massless particles (due to Klein tunneling).

The lack of a bandgap in graphene has attracted considerable attention from the researchers
in search of a mechanism for opening a gap between the $\pi$ and $\pi^*$ bands. In fact, the 
shape and size of the graphene flakes play a crucial role in defining its electronic properties 
\cite{nakada}. Small flakes have been found to possess a gap which is found to be suppressed 
and exhibit an oscillatory behavior with increasing width of the structures along the zigzag 
edges. A reduction of the gap with increasing size of the graphene flakes has indeed been 
confirmed experimentally \cite{han2007,ritter2009}. Occurrence of the gap was proposed to be 
induced by the quantum confinement effect in the flakes of small size. However, it was also 
shown that the specific ordering of the localized states at the zigzag edges of graphene 
(when localized states are ferromagnetically ordered along the zigzag edge and 
antiferromagnetically between the opposite zigzag edges) would break the sublattice symmetry 
of graphene and, therefore, can also open a gap \cite{pisani,son2006}. The oscillatory behavior 
has been attributed to the shape of the edges \cite{nakada}. 
 
\begin{figure}
\begin{center}
\includegraphics[scale=0.45]{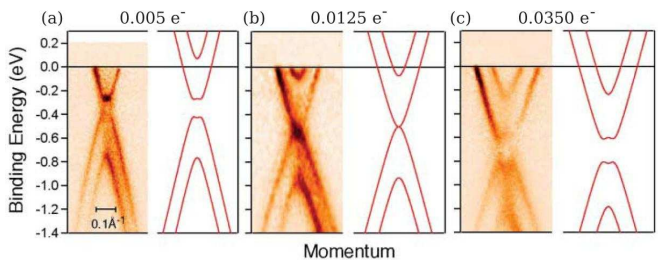}
\caption{\label{fig:fig2}
Experimental data on doping of bilayer graphene by adsorption of potassium \cite{ohta-sci313}.
Each panel is divided into two parts: the experimental results are on the left,
while the theoretical estimations obtained from tight-binding calculations are
on the right. Number of electrons per unit cell transferred from
potassium to graphene is indicated at the top of the each panel.}
\end{center}
\end{figure}

The idea of opening of an energy gap by breaking one of the symmetries in graphene 
(sublattice or lattice) has received wide recognition in the scientific community 
\cite{ohta-sci313,wehling2008,julia_ads}. It was found that any alteration of the 
ideal hexagonal lattice of graphene by defects would break the lattice symmetry of 
graphene, while the unequal charge exchange of different graphene sublattices with the 
substrate or adsorbates breaks the sublattice symmetry, thereby generating a gap. 
Experimentally, it was shown that graphene on a substrate indeed possesses a gap 
\cite{ohta-sci313}. It was also found that the charge exchange with the 
substrate or with the adsorbates would dope graphene and the type of doping ($p$- 
or $n$-type) can be controlled by the type of molecule that is adsorbed, i.e. the acceptor 
or the donor \cite{ohta-sci313,wehling2008}. The experimental data on alteration of 
the band structure of the bilayer graphene on SiC substrate with adsorption of potassium 
is displayed in Fig.~\ref{fig:fig2}. In as-prepared samples of graphene, the charge 
accumulated on the surface of graphene and the depleted SiC substrate create a 
built-in dipole field which leads to the opening of a gap through breaking of the 
graphene symmetry [see Fig.~\ref{fig:fig2} (a)]. Potassium adsorbed on the graphene 
surface dopes graphene and alters that dipole field, thereby modifying the size of 
the bandgap. When the number of transferred electrons per unit cell is 0.0125 \={e} 
the built-in dipole field is neutralized and the gap is closed (see Fig.~\ref{fig:fig2} 
(b)), while a further increase in concentration of the transferred electrons leads 
to reopening of the gap [see Fig.~\ref{fig:fig2} (c)]. Because potassium is a donor 
of electrons the charge exchange between potassium and graphene shifts the bands of 
graphene down according to the position of the Fermi level.

\section{Graphane -- hydrogenated graphene}

\begin{figure}
\includegraphics[scale=0.35]{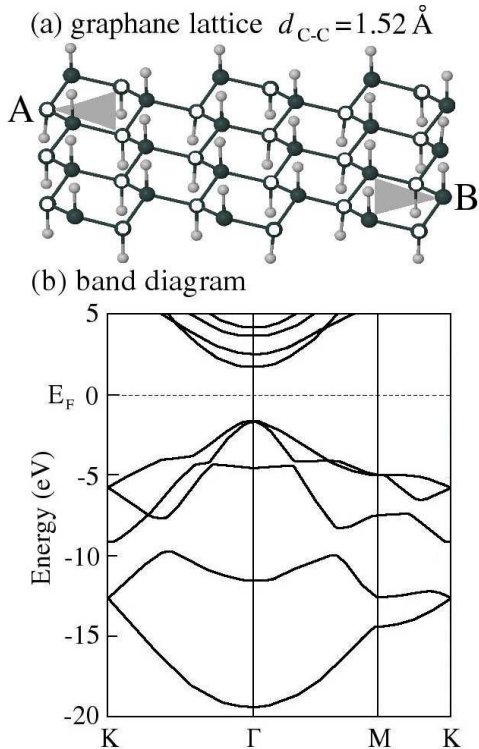}
\caption{\label{fig:fig3} (a) The graphane lattice in chair conformation (full and 
empty carbon atoms defines the A and B sublattices, respectively). (b) The band diagram 
of graphane \cite{sofo}. The Fermi level ($E_F$) is set to zero energy. The $\pi$ and 
$\pi^*$ bands are not present in the energy diagram of graphane because of the $sp^3$ 
hybridization of the carbon bonds in graphane.}
\end{figure}

The strong influence of molecular adsorption on the electronic properties of 
graphene has made this topic very popular among the researchers \cite{bouk,julia_ads}. 
Most adsorbed molecules interact weakly with pure graphene holding on to its surface 
or at the edges via the van der Waals forces. However, some adsorbates can bind to 
the carbon atoms in graphene which not only breaks the lattice symmetry 
but also modifies the $sp^2$ hybridization of the carbon bonds \cite{bouk}. For example, 
bonding of a carbon atom with adsorbate possessing one valence electron, such as H or 
Li \cite{sofo,yang,bouk}, leads to moving of this carbon atom out of the graphene plane 
and change the carbon bonds connecting this atom with its neighbors from the $sp^2$ 
hybridization to $sp^3$. In the case when all the carbon atoms are hydrogenated, 
the planarity of the graphene lattice is destroyed giving rise to a new lattice type which 
can adopt two main conformations: the so-called `boat' and `chair' conformations \cite{sofo}. 
In the boat conformation, hydrogenation of the carbon atoms occurs in pairs, i.e., pairs 
of nearest-neighbor carbon atoms are hydrogenated on the same side of the plane. 
However, the repulsion between two paired hydrogen atoms leads to instability of this 
conformation. The chair conformation is more stable because the nearest-neighbor carbon 
atoms, which belong to different sublattices, are hydrogenated from different sides of 
the graphene plane. The structure of the chair conformation is shown in Fig.~\ref{fig:fig3} (a). 
For this conformation, hydrogenation causes one sublattice to move out of the 
crystal plane, mimicking other $sp^3$-bonded crystal, such as diamond.
Therefore, the carbon atoms in ideal {\it graphane} (fully hydrogenated 
on both sides of a graphene lattice) lattice form two separate planes and for each plane 
the trigonal symmetry of the carbon atoms belonging to A- or B-sublattices is preserved.

%Figure  2

\begin{figure}
\includegraphics[scale=0.40]{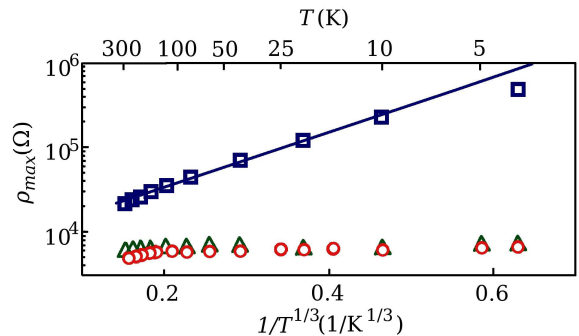}
\caption{\label{fig:fig4}
Temperature dependence of the resistivity of pristine graphene (circles), graphene 
after hydrogenation (squares) and hydrogenated graphene after annealing (triangles). 
The solid line is a fit obtained from the hopping dependence $\exp[(T_0/T)^{\frac13}]$, 
where $T_0$ is a parameter that depends on the gate voltage. From \cite{elias2009}.}
\end{figure}

The electronic structure of graphane was reported for the first time in theoretical work \cite{sofo}. 
The band diagram obtained there is presented in Fig.~\ref{fig:fig3} (b). 
All four valence electrons belonging 
to the carbon atoms participate in the formation of the covalent bonds and therefore 
the $\pi$ bands are removed from the band structure of graphane. As a result of the absence 
of the $\pi$ bands, which in graphene was responsible for the gapless nature of its 
electronic structure, graphane is a semiconductor characterized by a wide direct gap 
at the $\Gamma$ point \cite{sofo}. Moreover, transformation of the carbon bonds from the 
$sp^2$ to $sp^3$ hybridization results in an increase in the bond length from 1.42 \AA\ [see 
Fig.~\ref{fig:fig1} (a)] to 1.52 \AA\ [see Fig.~\ref{fig:fig3}(a)]. Subsequently, it 
was experimentally confirmed that hydrogenation of graphene alters its lattice structure 
resulting in the $sp^3$ hybridization of the carbon bonds \cite{ryu2008} that indeed 
alters the behavior of graphene in an electric field from conducting to insulating 
\cite{elias2009}. The resistivity of pristine graphene and hydrogenated graphene as a 
function of temperature are plotted in Fig.~\ref{fig:fig4}. Hydrogenation of graphene 
has been found to be reversible \cite{ryu2008,elias2009}, which is a definite advantage 
for application of graphene/graphane in nanoscale electronics. Moreover, the combination 
of pure and hydrogenated graphene distinguished by the different size of the bandgap 
is promising for creation of periodic multi-quantum arrays that facilitates the
resonant charge transfer \cite{chernoz2007}.  

\section{Magnetc properties of graphane}

In the original theoretical work where graphane was proposed \cite{sofo}, 
hydrogenation was performed from both sides of the graphene plane while in 
available experiments \cite{ryu2008,elias2009} graphene on a substrate was 
hydrogenated only from one side. Quite naturally, later theoretical work 
\cite{zhou} has focused on the electronic properties of free-standing graphene, 
hydrogenated from one side (which was christened {\it graphone}). It was claimed 
that graphone is a ferromagnetic semiconductor with a small indirect gap 
($E_g$=0.46 eV) \cite{zhou}. Ferromagnetism in graphone is attributed to the 
presence of localized states on the non-hydrated side. 

\begin{figure}
\includegraphics[scale=0.45]{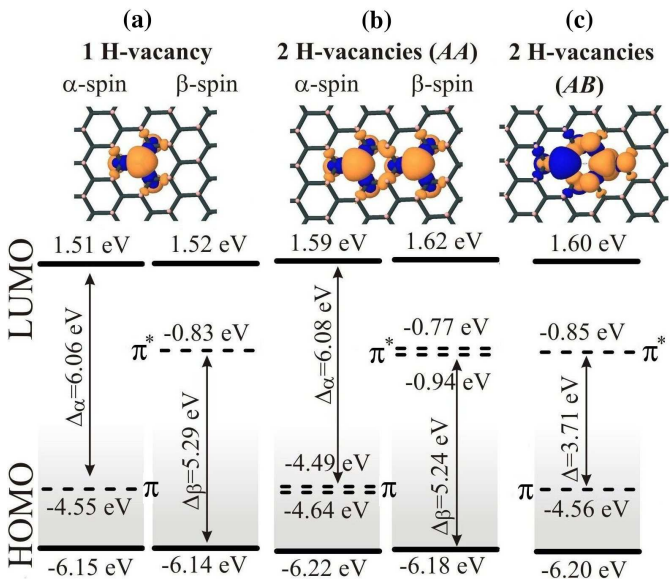}
\caption{\label{fig:fig5} Spin density (isovalues of $\pm 0.001$ e/\AA$^{3}$) 
and energetics of the bands in graphane with H-vacancies (the orbitals residing
on graphene are marked by the solid lines, while the defect levels $\pi$ and 
$\pi^*$ are marked by the dashed lines): (a) graphane containing a single 
H-vacancy; (b) graphane containing two H-vacancies located on one side of the 
graphane plane (AA-distribution) and separated by the distance of $d=4a^{}_{\rm 
C-C}$. (c) graphane containing two H-vacancies distributed between the two sides 
of the graphane plane (AB-distribution) and separated by a distance 
of $d=3a^{}_{\rm C-C}$ \cite{julia}.}
\end{figure}

In graphane each H-vacancy defect, 
which leaves an unsaturated dangling bond released on the carbon atom 
generates a localized state characterized by the 
unpaired spin. This spin-polarized state creates a defect level in the bandgap 
of graphane \cite{julia}. The band diagram of a finite size graphane (the size 
of the bandgap of defect-free graphane was $E_g$=7.51 eV in Ref.~\cite{julia}) 
containing a single H-vacancy defect is shown in Fig.~\ref{fig:fig5} (a). In the 
case of a single defect in graphane, the bandgap -- the energy difference between 
the lowest unoccupied (LUMO) and the highest occupied molecular orbitals (HOMO) 
-- is suppressed due to the appearance of the defect level between the HOMO and 
LUMO of the defect-free graphane. For the $\alpha$-spin state, the H-vacancy 
induces a defect level closer to the valence band which becomes the new HOMO, 
while for the $\beta$-spin state -- closer to the conduction band generating 
a new LUMO. As a result, the HOMO-LUMO gaps for the $\alpha$- and $\beta$-spin 
states are significantly shifted in energy.

%Figure  3

If there are two or more H-vacancies in graphane, the spins of the localized 
states can be ferromagnetically or anti-ferromagnetically ordered (see 
Fig.~\ref{fig:fig5} (b,c)). The ordering is defined by the distribution of 
the defects between the sides of the graphene plane, i.e., their actual 
presence in different sublattices. Therefore, if the two H-vacancy defects are 
located on the same side of the graphane plane (AA distribution) then the two 
localized states reside on the same sublattice. In graphene we already know that
for two states on the same sublattice if their spins are antiparallel then the 
destructive interference between the spin-up and spin-down tails of these states 
leads to an enhancement of the total energy of the system \cite{son}. Therefore, 
in both graphene and graphane, if the two states are localized on the same sublattice 
(AA distribution), the ferromagnetic ordering of their spins is energetically 
preferable. For the states localized on different sublattices (AB distribution), 
achieved in graphane by removing the hydrogen atoms from different sides of the 
graphane plane, antiferromagnetic ordering of the spins lowers the total 
energy of the system. Ferromagnetic and antiferromagnetic ordering of the 
spins versus the sublattice symmetry are in line with Lieb's theorem \cite{lieb}.

The destructive and constructive interference between the spin-up and spin-down 
tails of the localized states decrease exponentially with increasing distance 
between the states \cite{son}. For the AA distribution, ferromagnetic ordering 
of the spins is energetically preferable for $d\le4a^{}_{\rm C-C}$, otherwise, 
the interference of the tails is negligible and the difference of the total 
energies between the states characterized by different spin ordering is less 
than $\sim 10^{-2}$ eV \cite{julia}. Therefore, one can conclude \cite{julia} that 
the ferromagnetic ordering of spins of the localized states in graphane is possible 
only for the H-vacancy defects located on the neighboring carbon atoms. 

Ferromagnetic ordering of spins makes graphane containing the vacancy defects 
a ferromagnetic wide gap semiconductor. Because each vacancy can generate a magnetic 
moment of $\mu=1.0\mu^{}_B$, where $\mu^{}_B$ is the Bohr magneton, the magnitude 
of magnetism can be manipulated by the vacancy concentration ($N$). However, 
increasing vacancy concentration can make it difficult to maintain 
the ordering of the spins even when the vacancies are placed close to each other. 
Thus it was found that with an increasing number of H-vacancy defects, the state 
characterized by the highest magnetic moment gets closer in energy to several 
states of lower magnetic moment (when $\mu$ is less than the number of vacancies), 
thereby limiting the magnitude of maximum magnetization. Moreover, there is a 
critical number of defects $N\le$8 when the energy difference between the state of 
ferromagnetic  and antiferromagnetic ordering rapidly decreases (for $N$=8, $\Delta 
E$=6.56 eV while for $N$=10, $\Delta E$=0.25 eV), thereby destabilizing the ferromagnetic 
state. The lattice relaxation of graphane induced by the presence of defects also 
destabilizes the ferromagnetic state. Moreover, we believe that the non-uniform 
distribution of the hydrogen atoms over the graphane plane found in \cite{flores} 
(showing that the sequence of the up and down H atoms would be broken) 
would also have some issues in maintaining ferromagnetic ordering of the spins.

\section{Application in nanoscale electronics}

The unique electronic properties of both graphene and graphane makes them 
excellent candidates for nanoscale electronics. The influence of the 
substrate and the type of molecules adsorbed on the graphene surface on the 
electronic properties can be used to tune the size of the bandgap and the
doping in the production of nanoscale devices. It has, for example, 
already been applied in developing gas sensors with sensitivity at the
level of a single molecule \cite{wehling2008}. In graphane, the size of the gap 
can be manipulated by inducing the H-vacancy defects \cite{julia} or 
through the interaction of these vacancies with adsorbates.

The magnetic properties of graphene and graphane are also fascinating. 
Dependence of the ordering of spins of electrons in the localized states on 
their distribution between the sublattices is unique and suitable for
application of graphene/graphane in spintronics and magnetoelectronics.
Therefore, for graphene whose zigzag edges possess the localized 
states such that the spin-up and spin-down states are spatially 
separated between the opposite edges, half-metallicity of 
can be obtained by the application of an external electric field 
\cite{son2006}, by chemical functionalization or substitutional doping 
at the zigzag edges \cite{julia2}. Moreover, the imbalance in the distribution 
of the localized states between the sublattices (for example, by distribution 
of the defects predominantly on one sublattice) creates the non-zero 
magnetization in graphene/graphane.

The unique electronic and magnetic properties of graphene/graphane 
are expected to contribute significantly in future applications 
in nanoscale electronics. However, to make that a reality we must be able to control 
of shape and quality of the edges of graphene/graphane structures. 
In most of the methods developed for
fabrication of graphene, such as exfoliation of graphite \cite{nov2004},
lithographic patterning \cite{lit} and chemical sonication \cite{sonic}, 
the control of the flake size and quality of the edges are really poor. 
However, recently developed methods involving the unzipping of carbon nanotubes 
\cite{nature1,nature2} are very promising. It has been claimed that 
unzipping by plasma etching provides smooth edges and a small range of 
nanoribbon width (10-20 nm) \cite{nature1}. The second unzipping method 
-- the solution-based oxidative process -- allows one to perform the 
longitudinal cut of the nanotubes thereby creating flakes with 
predominantly straight linear edges \cite{nature2}. In graphane, for which 
the fluctuation of the shape and sizes of the flakes would not crucially influence 
its electronic properties because of the large bandgap, the main 
problem to be solved is a controllable dehydrogenation. The application of graphene 
and graphane in nanoscale electronics will have to wait until 
those technological issues are resolved satisfactorily. 

\section{Acknowledgements}
This work has been supported by the Canada Research Chair program and 
the NSERC Discovery Grant.

\end{document}